\documentclass[aps,prl,twocolumn,showpacs,preprintnumbers,amsmath,amssymb]{revtex4-1}

\usepackage{graphicx}
\usepackage{bm}
\usepackage{upgreek}

\begin{document}

\title{Crossover between short and long range proximity effects in SFS 
junctions with Ni-based ferromagnets.}

\author{O. M. Kapran$^1$, T. Golod$^1$, A. Iovan$^{1,2}$, A. S. Sidorenko$^{3,4}$, A. A. Golubov$^{5,6}$
and V. M. Krasnov $^{1,6}$} \email[E-mail:
]{Vladimir.Krasnov@fysik.su.se}

\affiliation{$^1$ Department of Physics, Stockholm University,
AlbaNova University Center, SE-10691 Stockholm, Sweden;}

\affiliation{$^2$ Department of Applied Physics, Royal Institute
of Technology, SE-10691 Stockholm, Sweden;}

\affiliation{$^3$ Institute of Electronic Engineering and
Nanotechnologies ASM, MD2028 Kishinev, Moldova; }

\affiliation{$^4$ I.S. Turgenev Orel State University, 302026
Orel, Russia;}

\affiliation{$^5$ Faculty of Science and Technology and MESA+
Institute of Nanotechnology, University of Twente 7500 AE,
Enschede, The Netherlands;}

\affiliation{$^5$ Moscow Institute of Physics and Technology,
State University, 141700 Dolgoprudny, Russia.}

\begin{abstract}
We study Superconductor/Ferromagnet/Superconductor junctions with
CuNi, PtNi,
or Ni interlayers. Remarkably, we observe that supercurrents
through Ni can be significantly larger than through diluted
alloys. The phenomenon is attributed to the dirtiness of
disordered alloys leading to a short coherence length despite a
small exchange energy. To the contrary, pure Ni is clean resulting
in a coherence length as long as in a normal metal.
Analysis of temperature dependencies of critical currents reveals
a crossover from short (dirty) to long (clean) range proximity
effects in Pt$_{1-x}$Ni$_x$ with increasing Ni concentration. Our
results point out that structural properties of a ferromagnet play
a crucial role for the proximity effect and indicate that
conventional strong-but-clean ferromagnets can be advantageously
used in superconducting spintronic devices.

\end{abstract}



\maketitle

\section{I. Introduction}

A competition between superconductivity and ferromagnetism leads
to an unconventional proximity effect, studied both theoretically
\cite{Beenakker_1995,Demler_1997,Kadigrobov_2001,Tagirov_2003,Eschrig_2004,Buzdin2005,Efetov2005,Radovic_2005,Golubov,Silaev_2009,Fominov,Buzdin_2010,Aladoust_2010,Pugach_2011,Melnikov_2012,Linder_2015,Silaev_2017,Klenov_2019}
and experimentally
\cite{Petrashov_1999,Ryazanov_2001,Chandrasekhar_2001,Kontos_2002,Oboznov_2006,Born_2006,Keizer_2006,Robinson_2007,Bannykh_2009,Robinson_2010,Aarts_2010,Wang_2010,Birge_2011,Bolginov_2012,Golod_2013a,Dresselhaus_2014,Kompaniits_2014,Iovan_2014,Sidorenko_2016,Aarts_2017,Birge_2018,Skryabina_2019,Kapran_2020}.
In strong ferromagnets (F) Fe, Co, Ni, exchange energies,
$E_{ex}\sim 1000$ K, are much larger than the energy gap, $\Delta
\sim 1-10$ K, in low-$T_c$ superconductors (S). Therefore,
spin-singlet Cooper pairs are usually broken
at a very short range $\sim 1$ nm in F, as shown by many
experimental works
\cite{Chandrasekhar_2001,Robinson_2007,Bannykh_2009,Robinson_2010,Birge_2011,Dresselhaus_2014,Birge_2018,Skryabina_2019}.
There are, however, reports about a long range proximity effect
(LRPE) (tens to hundreds of nm)
\cite{Petrashov_1999,Keizer_2006,Wang_2010,Golod_2013a,Kompaniits_2014},
which is often ascribed to the spin-triplet order parameter that
should be immune to the ferromagnetic order.

Interpretation of LRPE remains controversial. First, there is a
seeming irreproducibility of experimental results,
cf. Refs. \cite{Petrashov_1999,Wang_2010,Kompaniits_2014} and
\cite{Chandrasekhar_2001,Skryabina_2019}. Second, the triplet
order should appear only in the noncollinear magnetic state
\cite{Buzdin2005,Efetov2005,Fominov}, the origin of which is often
unclear for structures containing only one F layer. Although
several subtle effects, such as quantum fluctuations
\cite{Kadigrobov_2001}, active interfaces \cite{Eschrig_2004},
domains \cite{Robinson_2010,Aarts_2017}, inhomogeneities
\cite{Silaev_2009,Aladoust_2010,Silaev_2017} and spin-orbit
coupling \cite{Linder_2015,Silaev_2017} were suggested,
they are difficult to confirm or control in experiment. Finally,
the proximity effect depends on the electronic mean-free path
(m.f.p.), $l_e$, and, thus, on the internal structure. In
particular, it has been predicted, that in clean F even a singlet
supercurrent should exhibit LRPE
\cite{Demler_1997,Radovic_2005,Buzdin_2010,Pugach_2011,Melnikov_2012,Born_2006}.
Experimental analysis of SFFS spin-valves
has shown that the singlet current is dominant for diluted F
\cite{Iovan_2014} and remain considerable even for pure Ni
\cite{Kapran_2020}. Clarification of LRPE mechanisms and the ways
of controlling supercurrents in S/F heterostructures is important
both for fundamental understanding of unconventional
superconductivity \cite{Balatsky_2019}, and for application in
superconducting spintronics
\cite{Bolginov_2012,Dresselhaus_2014,Birge_2018,Klenov_2019,Kapran_2020}.

\begin{figure*}[th]
    \centering
    \includegraphics[width=0.99\textwidth]{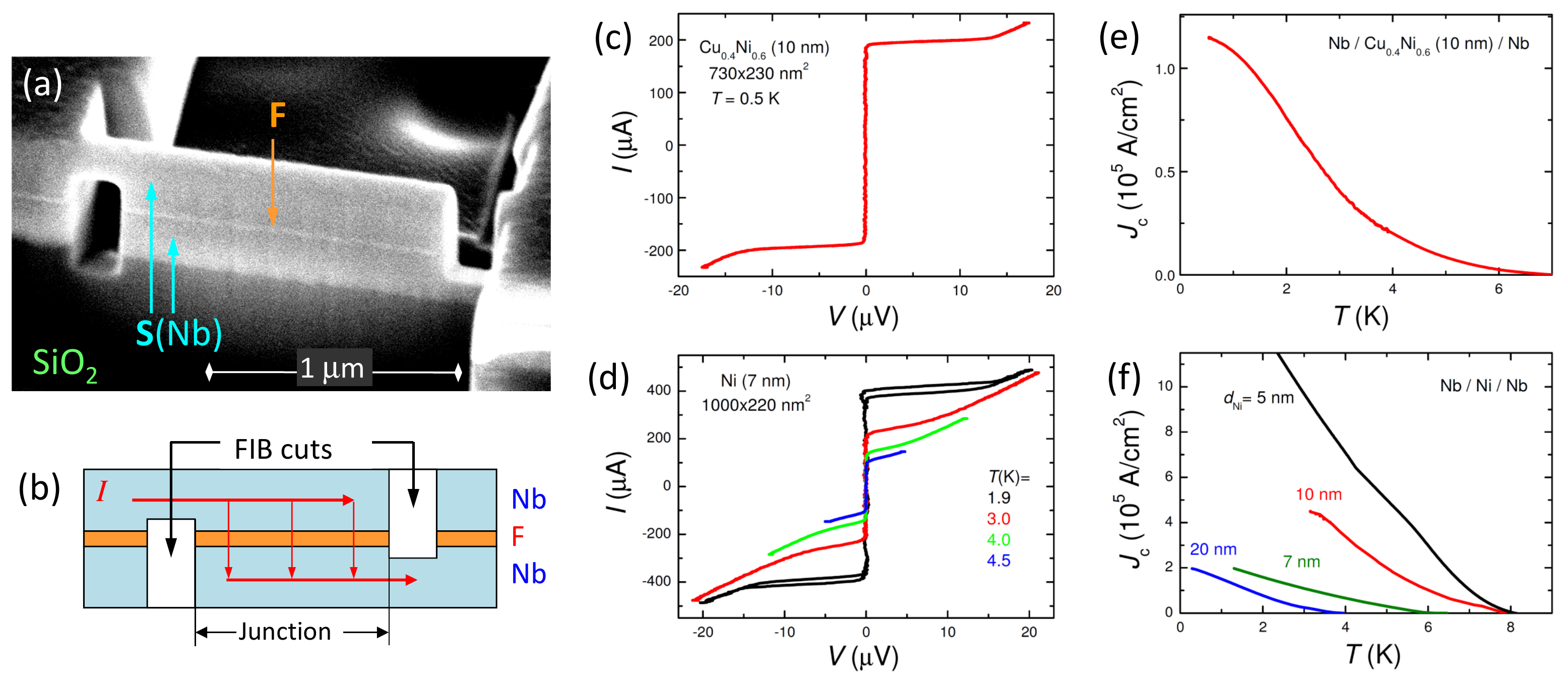}
    \caption{(color online). (a) SEM image of an SFS junction and (b) a sketch with indication of current flow paths. (c) The $I$-$V$ curve of Nb/Cu$_{0.4}$Ni$_{0.6}$/Nb junction with $d_F=10$ nm.
    (d) A set of $I$-$V$'s of a Nb/Ni(7 nm) /Nb junction at different $T$. (e, f ) Temperature dependencies of the critical current density for (e) the same Nb/Cu$_{0.4}$Ni$_{0.6}$/Nb junction
    and (f) Nb/Ni/Nb junctions with different Ni thicknesses 5, 7, 10 and 20
    nm. Note that the $J_c(T=3~K)$ for the Ni-junction, $d_{F}=10$ nm, is almost an order of magnitude larger than that
    for the junction with diluted Cu$_{0.4}$Ni$_{0.6}$ interlayer with the same $d_F$.
} \label{fig:fig1}
\end{figure*}

Here we study nanoscale SFS Josephson junctions (JJ's) containing
either diluted Ni-alloys Cu$_{1-x}$Ni$_x$ and Pt$_{1-x}$Ni$_x$,
Cu/Ni bilayer or pure Ni.
Counterintuitively, we observe that the supercurrent density,
$J_c$, through Ni can be much larger than through diluted alloys
with the same thickness.
Using {\em in-situ} absolute Josephson fluxometry (AJF), we
demonstrate that Ni interlayers in our junctions exhibit full
saturation magnetization as in bulk Ni, which precludes presence
of extended dead magnetic layers. The clue to understanding of our
results is obtained from the analysis of evolution of temperature
dependencies, $J_c(T)$, in Nb/Pt$_{1-x}$Ni$_x$/Nb JJ's with
increasing Ni concentration. It shows that in diluted Ni-alloys,
$x\simeq 0.5$, the proximity effect is short range, despite a
small $E_{ex}$, due to an extremely short m.f.p. is such
atomically disordered alloys. To the contrary, pure Ni remains
clean, facilitating ballistic Cooper pair transport and LRPE
similar in scale to that in the normal metal Pt. Our results
demonstrate that the proximity effect in ferromagnets depends not
only on composition and $E_{ex}$, but also essentially on the
internal structure. This may help to resolve some of the
controversies around LRPE. We conclude that strong-but-clean
ferromagnets may have advantages compared to weak-but-dirty for
device applications.

The paper is organized as follows. In sec. II we describe sample
fabrication and experimental procedures. In sec. III we discuss
main experimental results, including III A. {\em in-situ} magnetic
characterization of Ni interlayers via AJF and III B. analysis of
temperature dependencies of critical currents, which reveals a
crossover between clean (ballistic) and dirty (diffusive)
transport. In the Appendix we provide additional information about
A. film structure, B. junction characteristics, C. properties of
Nb/Pt$_{1-x}$Ni$_x$/Nb junctions, D. interface resistances in
Nb/Pt$_{1-x}$Ni$_x$/Nb junctions, and E. extraction of
magnetization curves from AJF analysis.

\section{II. Samples and Experimental}

We present data for nano-scale SFS junctions with F-interlayers
made of Cu$_{0.4}$Ni$_{0.6}$ and Pt$_{1-x}$Ni$_x$ alloys with
$x=0-1$, pure Ni and a Cu/Ni (N/F, N-normal metal) bilayer. SFS
multilayers were deposited by dc-magnetron sputtering in a single
cycle without breaking vacuum. Cu$_{1-x}$Ni$_x$ films were
deposited by cosputtering from Cu and Ni targets and the
concentration was controlled by the corresponding sputtering
rates. Pt$_{1-x}$Ni$_x$ films were deposited from composite
targets with different areas of Ni and Pt segments and Ni
concentration was estimated using energy-dispersive X-ray
spectroscopy. More details about fabrication and magnetic
properties of Pt$_{1-x}$Ni$_x$ films can be found in Refs.
\cite{Golod_2011,Golod_2013b} and in Appendices C and D. Nb/Ni/Nb
JJ's with different $d_F$ were fabricated from the same wafer with
a calibrated Ni-thickness gradient \cite{Born_2006}. Nb(S)
electrodes were $\sim 200$ nm thick. Multilayers were first
patterned by photolithography and reactive ion etching and then
processed by focused ion beam (FIB). Nano-scale JJ's with sizes
down to $\sim 60$ nm were made by FIB-nanosculpturing
\cite{Robinson_2007,Golod_2010,Iovan_2014}. Small sizes are
necessary both
for achieving the monodomain 
state \cite{Iovan_2017,Kapran_2020} and for enhancing normal
resistances to comfortably measurable values, $R_n \gtrsim 0.1
\Omega$. We present data for JJ's with different sizes, interlayer
thicknesses, $d_F$, and compositions. Junction parameters are
listed in Tables I-III of the Appendix. Properties of Josephson
spin valves with similar CuNi and Ni interlayers can be found in
Refs. \cite{Iovan_2014} and \cite{Kapran_2020}. Figure
\ref{fig:fig1} shows (a) a scanning electron microscope (SEM)
image of one of the studied Nb/Ni/Nb JJ's and (b) a sketch with a
current path.

Measurements are performed in $^3$He 
and $^4$He 
closed-cycle cryostats. Magnetic field, parallel to the junction
plane, is supplied by a superconducting solenoid. We will show
measurements with field oriented either parallel $H_{\parallel}$
(easy axis), or perpendicular $H_{\perp}$ (hard axis) to the long
side of the JJ.

\begin{figure*}[t]
    \centering
    \includegraphics[width=0.99\textwidth]{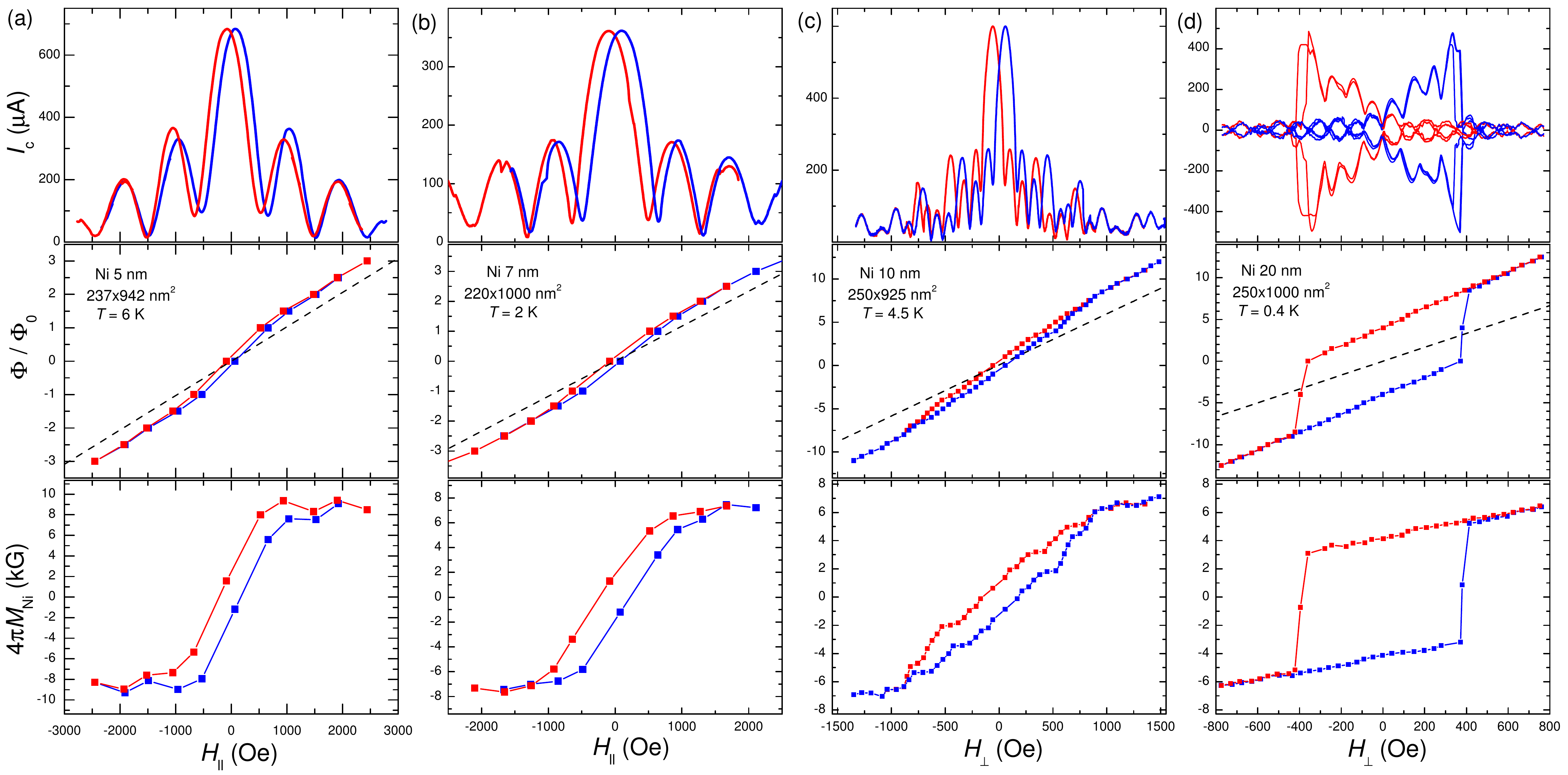}
    \caption{(color online). Top panels: magnetic field modulation of the critical current, $I_c(H)$, for Nb/Ni/Nb JJ's
    with $d_{Ni}$ (a) 5 nm and (b) 7 nm in the easy axis orientation and (c) 10 nm and (d) 20 nm in the hard axis orientation.
    Blue/red lines represent up/down field sweeps. Middle panels show the absolute Josephson fluxometry analysis of the data above.
    Symbols represent positions of maxima and minima of the $I_c(H)$ patterns, which correspond to half-integer and integer values of $\Phi/\Phi_0$.
    Bottom panels represent magnetization curves of Ni interlayers obtained from the AJF analysis. Large values of the saturation magnetization $4\pi M_{Ni} \simeq 7$ kG
    are consistent with the value for bulk Ni and preclude presence of dead magnetic layers.
}
    \label{fig:fig2}
\end{figure*}

\section{III. Results and discussion}

Figs. \ref{fig:fig1} (c,d) show Current-Voltage characteristics
($I$-$V$) at zero field for (c) Nb/Cu$_{0.4}$Ni$_{0.6}$/Nb
junction with $d_F=10$ nm at $T\simeq 0.5$ K and (d) Nb/Ni/Nb
junction with $d_F=7$ nm at different temperatures, $T=1.9-4.5$ K.
The shapes of $I$-$V$'s are typical for proximity-coupled JJ's,
described by the resistively shunted junction model.

Figs. \ref{fig:fig1} (e,f) show temperature dependencies of
critical current densities for (e) the same
Nb/Cu$_{0.4}$Ni$_{0.6}$/Nb JJ, and (f) Nb/Ni/Nb JJ's with $d_F=$
5, 7, 10 and 20 nm. It is seen that the JJ with a diluted
Cu$_{0.4}$Ni$_{0.6}$ interlayer has a significantly smaller $J_c$
than the JJ with pure Ni with the same $d_F=10$ nm, compare red
lines in Figs. \ref{fig:fig1} (e) and (f). It is also seen that
the Cu$_{0.4}$Ni$_{0.6}$ JJ exhibits stronger superlinear
temperature dependence 
with a positive curvature $d^2 J_c/dT^2>0$ at elevated $T$, which
is well described by the power-law dependence $J_c\propto
(1-T/T_c)^a$ with $a\simeq 3.5$.
On the other hand, Ni JJ's show almost linear $J_c(T)$,
irrespective of Ni thickness, albeit with a varying onset
temperature $T_c^*$.

\subsection{III A. {\em In-situ} magnetic characterization of Ni interlayers via absolute Josephson fluxometry.}

Top panels in Figure \ref{fig:fig2} represent measured $I_c(H)$
modulation patterns for Nb/Ni/Nb JJ's with different $d_{Ni}$ (a)
5 nm, (b) 7 nm, (c) 10 nm and (d) 20 nm. Junction sizes and
measurement temperatures are indicated in the Figure. Modulation
patterns are shown both for easy (a,b) and hard (c,d)
axis orientations. 
Blue and red lines represent up and down field sweeps. A
hysteresis is due to finite coercivity of F-interlayers. It
disappears at $H \sim \pm (1-1.5)$ kOe, corresponding to
transition into the saturated magnetic state. All JJ's, included
in the analysis, exhibit Fraunhofer-type $I_c(H)$ modulation,
indicating good uniformity of interlayers \cite{Krasnov_1997}.
Examples of $I_c(H)$ patterns for Nb/PtNi/Nb and Nb/Ni/Nb
junctions can be found in Refs. \cite{Golod_2010} and
\cite{Kapran_2020}, respectively.

The $I_c(H)$ modulation occurs due to flux quantization. This can
be used for {\em in-situ} AJF analysis
\cite{Bolginov_2012,Iovan_2014,Kapran_2020}, presented in middle
panels of Fig. \ref{fig:fig2}. Here symbols represent the flux,
$\Phi (H)$, at maxima and minima of $I_c(H)$, which correspond to
half-integer and integer number of the flux quantum, $\Phi_0$,
respectively. The total flux is:
\begin{equation}\label{F_H}
\Phi= B L \Lambda + 4\pi M_F L d_F,
\end{equation}
where $B$ is magnetic induction, $L$ is the junction length,
$\Lambda$
is the effective magnetic thickness of the JJ
and $M_F$ is magnetization of the F-layer along the field. The
first term in the right-hand-side represents the flux induced by
magnetic field, the second - by magnetization of the F-layer (for
more details see Appendix E).

From Figs. \ref{fig:fig2} (a,b) it is seen that at large fields
$\Phi(H)$ is linear. Since in this case F-layers are in the
saturated state, $M_F=M_{sat}$, the linear field dependence is
caused solely by the first term in Eq. (\ref{F_H}) with $B\propto
H$. Subtraction of this linear dependence, shown by dashed lines
in middle panels of Fig. \ref{fig:fig2}, reveals the contribution,
$\Delta \Phi$, from the second term in Eq. (\ref{F_H}). This
yields the absolute value of magnetization in the F-interlayer
$4\pi M_F=\Delta\Phi/L d_F$. Thus obtained magnetization curves,
$4\pi M_{Ni}(H)$, are shown in bottom panels of Fig.
\ref{fig:fig2}. Saturation magnetizations are $4\pi M_{sat}=8.2\pm
1.1$ kG for $d_{Ni}=5$ nm, $7.3 \pm 0.5$ kG for $d_{Ni}=7$ nm and
$6.9 \pm 0.3$ kG for $d_{Ni}=10$ nm. For $d_{Ni}=20$ nm the
saturation magnetization is not reached within the shown field
range (see Appendix E for clarifications).
The main uncertainty in $M_{sat}$ is caused by the accuracy of
estimation of $d_{Ni}$, limited by the film roughness $R_q\simeq
1$ nm (see Appendix A). The thinner is the film - the larger is
such systematic uncertainty.

The obtained saturation magnetization $4\pi M_{sat} \simeq 7$ kG
is consistent with 
that for bulk nickel \cite{Danan_1968,Aldred_1975}. This implies
that Ni interlayers in our JJ's are fully ferromagnetic and there
are no extended dead magnetic layers, i.e., interface layers of Ni
with reduced magnetism. Such dead layers, accompanied by a
significant reduction of $M_{sat}$, were reported in earlier works
\cite{Oboznov_2006,Robinson_2007} and would make interpretation of
proximity effect more complicated.
On the other hand, a variation of the superconducting onset
temperature, which can be seen in Figs. \ref{fig:fig1} (f) and
\ref{fig:fig3} (b), provides an evidence for existence of dead
superconducting (rather than magnetic) layers with suppressed
$T_c^*$ at junction interfaces.

The large value of $M_{sat}$ confirms that supercurrents in our
Nb/Ni/Nb JJ's flow through a pure Ni with strong ferromagnetic
properties. Remarkably, we observe a large $J_c \simeq 2\times
10^5$ A/cm$^2$, even through 20 nm of Ni, see Fig. \ref{fig:fig1}
(f). This is a much longer scale compared to earlier reports
\cite{Chandrasekhar_2001,Robinson_2007,Bannykh_2009,Birge_2011,Dresselhaus_2014}
in which supercurrent was observed only through few nm of Ni.
Observation and clarification of such a profound LRPE through a
strong F is the main objective of this work.


\subsection{III B. Temperature dependencies of critical currents: crossover between dirty and clean regimes.}

To clarify our observations, we start with a short summary of
proximity effects in SNS and SFS JJ's (more detailed analysis can
be found in Ref. \cite{RevModPhys2004}, where various regimes have
been considered). In SNS JJ's $J_c$ is determined by the
superconducting order parameter at the junction interface,
$\Psi_{S/N}$, and the ratio of the thickness, $d_N$, to the
coherence length, $\xi_N$, of the interlayer. Close to $T_c$ it
can be written in the following simple form \cite{Golubov_1992},
\begin{equation}\label{Ic_T}
J_c\propto \Psi_{S/N}^2\exp\left(-\frac{d_N}{\xi_N}\right).
\end{equation}
For JJ's with thick $d_N$, or short $\xi_N$, $d_N\gg\xi_N(T_c)$,
the $J_c(T)$ is determined predominantly by the $T$-dependence of
$\xi_N(T)$, leading to a strong superlinear $T$-dependence. In the
opposite case, $d_N\ll\xi_N(T_c)$, $J_c(T)$ is determined by
$\Psi_{S/N}(T)$, leading to a conventional linear $J_c(T)$ close
to $T_c$ and a saturation at $T\rightarrow 0$.

For SFS JJ's the coherence length $\xi_F$ is complex
\cite{Buzdin2005,Ryazanov_2001},
\begin{equation}\label{Ksi_F}
\xi_F^{-1}=\xi_{F1}^{-1}+i\xi_{F2}^{-1}.
\end{equation}
The real part, $\xi_{F1}$, represents the decay length, the
imaginary, $\pi\xi_{F2}$, the period of oscillations. In the clean
case,
\begin{equation}\label{Ksi_Fclean}
\xi_F(c)=\frac{\hbar v_{f}}{2(\pi k_B T + i E_{ex})},
\end{equation}
where $v_f$ is the Fermi velocity in F.
In the dirty case,
\begin{equation}\label{Ksi_Fdirty}
\xi_F(d)=\sqrt{\frac{l_e}{3} \xi_F(c)}.
\end{equation}
Since here $l_e\ll|\xi_F(c)|$, the coherence length in the dirty
case is shorter than in the clean case. For strong F with
$E_{ex}/k_B \gg T_c$,
in the dirty case
$\xi_{F1}(d)\simeq \xi_{F2}(d)\simeq (\hbar l_e v_f / 3
E_{ex})^{1/2}$ are equaly short. However, in the clean case
the two scales are different: $\xi_{F1}(c)\simeq (\hbar v_f / 2\pi
k_B T)$, is as long as $\xi_N$, and $\xi_{F2}(c)\simeq (\hbar v_f
/ 2E_{ex})$ is short
\cite{Radovic_2005,Pugach_2011,Melnikov_2012,Born_2006}. From the
discussion above, it follows that the shape of $J_c(T)$ provides
an important clue about the proximity effect
\cite{Ryazanov_2001,Kontos_2002,Oboznov_2006,Robinson_2007}.

\begin{figure}[t]
    \centering
    \includegraphics[width=0.4\textwidth]{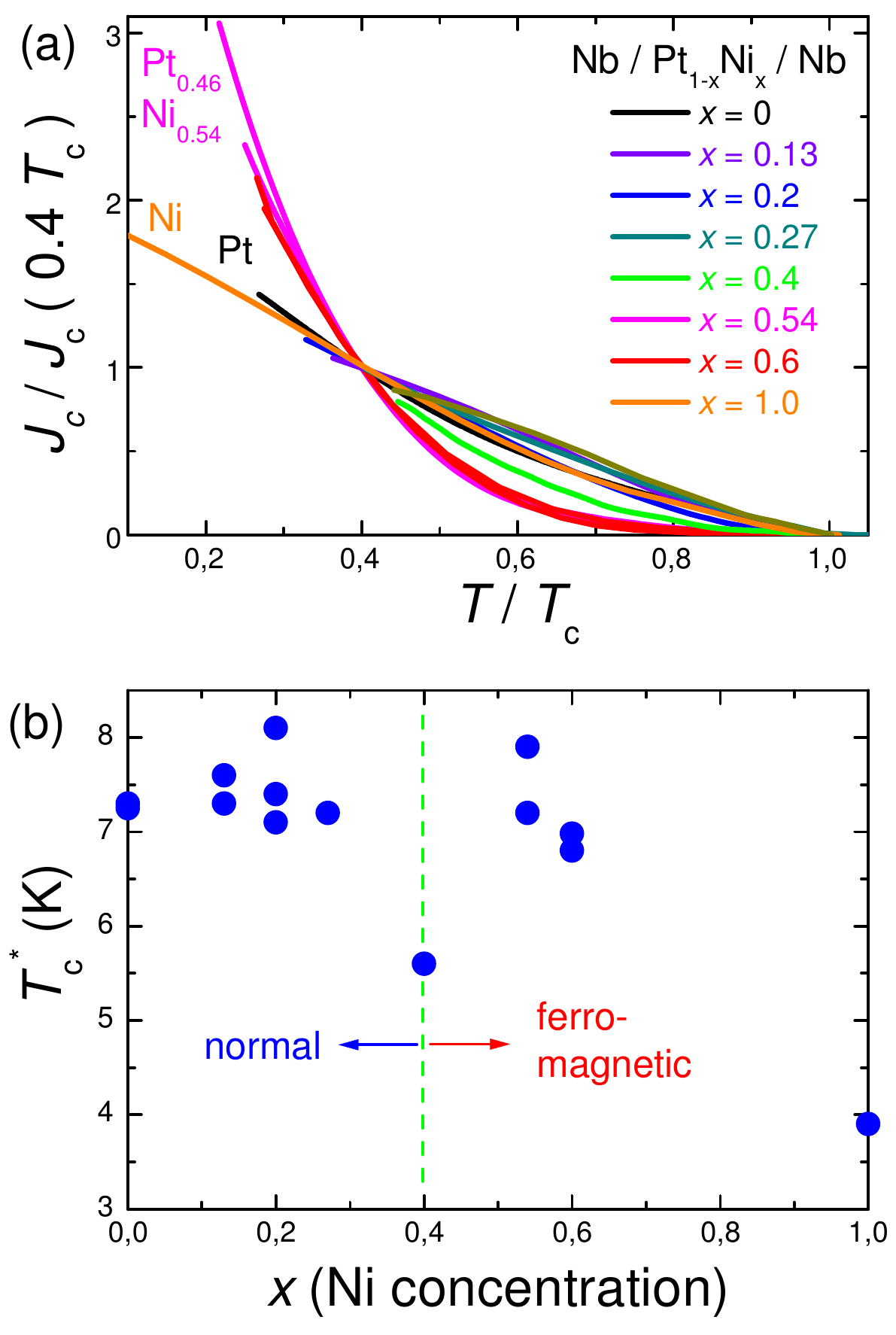}
    \caption{(color online). (a) Temperature dependencies $J_c(T)$ of Nb/Pt$_{1-x}$Ni$_{x}$/Nb junctions with $d_F=20$
    nm, normalized to the value at $T=0.4$ of the onset
    temperature. A gradual evolution of $J_c(T)$ dependencies from linear for pure Pt, $x=0$, to superlinear at $x\simeq 0.5-0.6$ and
    back to linear for pure Ni, $x=1$, indicates a transitions between SNS to dirty SFS and to clean SFS cases. (b) Corresponding onset
    temperatures of the junctions versus Ni concentration. 
}
    \label{fig:fig3}
\end{figure}

Figure \ref{fig:fig3} (a) shows evolution of $J_c(T)$, normalized
to the value at $0.4 T_c^*$ for Nb/Pt$_{1-x}$Ni$_x$/Nb JJ's with
different Ni
concentrations 
and $d_F\simeq 20$ nm. The onset temperature $T_c^*$ is shown in
Fig. \ref{fig:fig3} (b). Ferromagnetism in Pt$_{1-x}$Ni$_x$
appears at a critical concentration $x_c=0.4$
\cite{Golod_2011,Golod_2013b}, as described in the Appendix C. In
Fig. \ref{fig:fig3} (b), apart from a minimum of $T_c^*$ for the
pure Ni, $x=1$, we can also see a clear minimum at $x_c=0.4$. Both
minima can be interpreted as being due to suppression of
$\Psi_{S/N}$ at the interface (dead superconducting layer) due to
either a reverse magnetic proximity effect for pure Ni, $x=1$, or
quantum fluctuations at the quantum critical point, $x_c=0.4$
\cite{Golod_2013b}.

From Fig. \ref{fig:fig3} (a) it can be seen that for a pure Pt,
$x=0$ (black line), $J_c(T)$ is almost linear. As explained above,
this is expected for SNS JJ's with $\xi_N(T_c)\gg d_N$. Upon
increasing Ni concentration, $J_c(T)$ remains linear for
non-magnetic interlayers $x=0.13$, 0.2, 0.27. At the critical
concentration, $x_c=0.4$ (green line), a positive curvature
develops in $J_c(T)$. It becomes most pronounced for $x=0.5-0.6$
(magenta and red curves). Such a transformation is the consequence
of a rapid reduction of $\xi_F$ both due to enhancement of
$E_{ex}$ and reduction of $l_e$. While $E_{ex}$ increases linearly
at $x>x_c$, the m.f.p. reaches minimum at $x\simeq 0.5$,
corresponding to atomically disordered, $l_e\sim 1$ nm,
\cite{Pugach_2011} dirty metal (see Appendix C). With further
increase of concentration to pure Ni, $x=1$ (orange line), the
linear $J_c(T)$ dependence is restored, similar to a pure Pt,
$x=0$. Such a recovery implies that $\xi_{F1}$ in pure Ni is
similarly long as $\xi_N$ for pure Pt, despite the large $E_{ex}
\sim 10^3$ K. As discussed above, this indicates occurrence of
clean, ballistic transport in Ni. Thus, variation of the shape of
$J_c(T)$ reveals two crossovers in electron transport regimes with
changing Ni-concentration. First a crossover takes place from a
clean SNS type proximity effect in pure Pt to dirty SFS case in
diluted, atomically disordered alloy $x\simeq 0.5$. With further
increase of $x$ a second crossover from dirty to clean ballistic
transport takes place for JJ's with pure Ni.

As mentioned in the Introduction, interpretation of LRPE in strong
ferromagnets is still controversial. In several cases it was
attributed to appearance of the unconventional odd-frequency
spin-triplet order parameter. Recently the dominant ($\sim 70\%$)
spin-triplet supercurrent was reported in SF$_1$F$_2$S Josephson
spin-valve structures with similar Ni-interlayers
\cite{Kapran_2020}. However, the triplet supercurrent appears only
in the noncollinear state of the spin valve and is tuned by the
relative orientation of magnetization in the two F-layers.
Appearance and disappearance of the long-range triplet
supercurrent upon remagnetization of the spin-valve leads to a
profound distortion of the $I_c(H)$ pattern \cite{Kapran_2020}.
Such a distortion is the main fingerprint of the triplet component
\cite{Iovan_2017} and, thus, provides the key evidence for it's
existence. SFS junctions, containing just a single F-layer, behave
completely differently (see e.g. the discussion in sec. IV C of
Ref. \cite{Kapran_2020}). In particular, $I_c(H)$ patterns of all
our junctions are Fraunhofer-like, with the only distortion caused
by the hysteresis in $M_F(H)$. As discussed in the Introduction,
the triplet state is not anticipated in SFS junctions because
there is no obvious mechanism for appearance of the noncollinear
magnetic state in the perpendicular direction across the single
F-layer. Therefore, we want to emphasize, that LRPE in Nb/Ni/Nb
JJ's with clean Ni is achieved by the spin-singlet current without
involvement of the unconventional odd-frequency spin-triplet order
parameter. Such LRPE is simply a consequence of the lack of
scattering mechanism that can destroy singlet Cooper pair
correlations in a clean metal (no matter F or N) at $T=0$
\cite{Buzdin_2010}. Thus, it is the cleanliness of pure Ni that
facilitates LRPE in Nb/Ni/Nb JJ's. Concurrently, the extreme
dirtiness suppresses proximity effect through diluted F-alloys,
despite a small $E_{ex}$.

We also studied Nb/Cu(10nm)/Ni(10nm)/Nb JJ's, with Cu/Ni bilayer.
Interestingly, they show an order of
magnitude smaller $J_c$ 
than Nb/Ni(10nm)/Nb JJ's, see Table I in the Appendix, consistent
with earlier results for Ni-based JJ's with Cu buffer layers
\cite{Bannykh_2009,Birge_2011,Dresselhaus_2014}. This is
surprising because, due to a large $\xi_N\sim 1~\mu$m of Cu, 10 nm
should have little influence. On the other hand, neighbors in the
periodic table Cu and Ni tend to easily alloy with each other.
Therefore, Cu/Ni bilayers likely contain a dirty CuNi interlayer,
which leads to suppression of $J_c$.

\section{Conclusions}
To conclude, we have studied SFS junctions with different Ni-based
interlayers. We observed that supercurrents through pure Ni may be
much larger than through diluted alloys with much smaller
$E_{ex}$.
Analysis of $J_c(T)$ dependencies revealed that this
counterintuitive result is caused by the dirtiness of disordered
Ni-alloys, leading to a short coherence lengths $\xi_{F1}\sim 1$
nm. To the contrary, the mean-free-path in pure Ni interlayers can
easily exceed the film thickness \cite{Note2} up to several tens
of nm, facilitating ballistic Cooper pair transport with the decay
length as long as in non-magnetic normal metals. Our observation
suggests that SFS junctions with strong-but-clean ferromagnets may
have significant advantages, compared to commonly considered
weak-but-dirty alloys.

Our results may also help to resolve the controversy around LRPE
in strong ferromagnets, which is either seen
\cite{Petrashov_1999,Wang_2010,Kompaniits_2014} or not
\cite{Chandrasekhar_2001,Skryabina_2019}. We want to emphasize
that proximity effect in ferromagnets essentially depends on the
internal structure. In contrast to SNS JJ's, which always show
LRPE at low enough temperatures because $\xi_N(T\rightarrow
0)\rightarrow \infty$ irrespective of cleanliness,
for SFS JJ's LRPE occurs only in the clean case, for which
$\xi_{F1}(c)\rightarrow\infty$ at $T\rightarrow 0$, while for the
dirty case $\xi_{F1}(d)$ remains short irrespective of $T$. This
leads to a principle difference in the range of proximity effects
for clean and dirty ferromagnets with otherwise similar
compositions and exchange energies.


\begin{acknowledgments}
The work was supported by the EU H2020-WIDESPREAD-05-2017-Twinning
project ``SPINTECH", grant agreement Nr. 810144 (sample
preparation and measurements) and the Russian Science Foundation
grant No. 19-19-00594 (data analysis and manuscript preparation).
The manuscript was written during a sabbatical semester of V.M.K.
at MIPT, supported by the Faculty of Natural Sciences at SU.
\end{acknowledgments}

\begin{figure*}[t]
    \centering
    \includegraphics[width=0.7\textwidth]{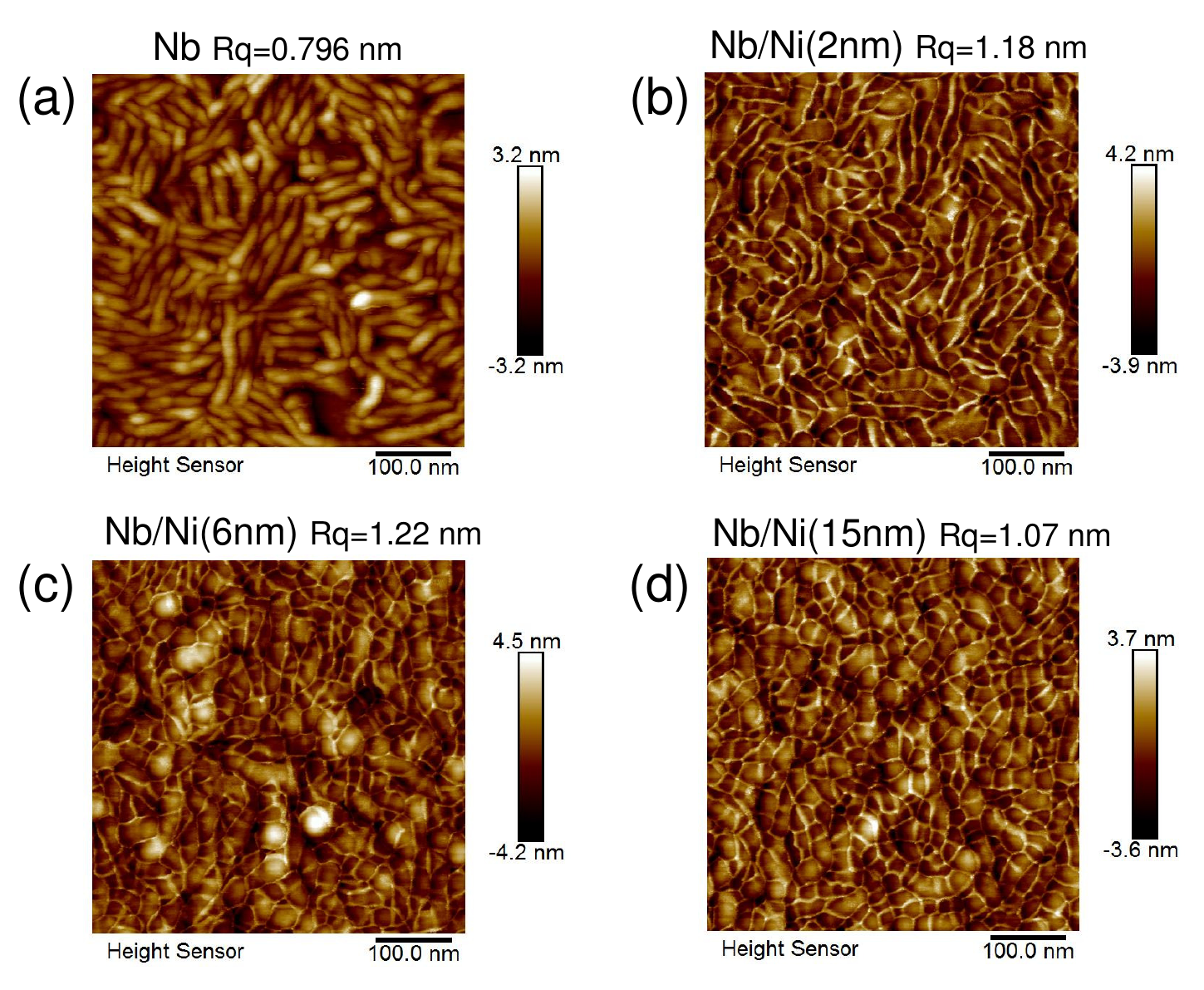}
    \caption{(color online). Atomic force microscope topography maps of a 100 nm thick Nb film (a) without Ni on
    top, and (b) with 2 nm Ni, (c) 6 nm Ni and (d) 15 nm Ni films on top. It can be seen that the Ni film reconstruction occurs at about 5 nm thickness.
} \label{fig:figS1}
\end{figure*}

\subsection{Appendix A. Nb/Ni film structure}

Figure \ref{fig:figS1} shows topography maps obtained by atomic
force microscopy for (a) a Nb film with thickness $d=100$ nm and
(b-d) Nb/Ni bilayers with increasing Ni thickness. It can be seen
that the Nb film has a rise-seed-like structure with elongated
crystallites (a). In Nb/Ni bilayers, with increasing Ni thickness,
$d_{Ni}$, the structure of Ni first inherits that of Nb (b) but at
$d_{Ni}\simeq 5$ nm a reconstruction to square-shaped crystallites
occurs (c), which do not change significantly in shape and size (
$\sim 20$ nm) with further increase of $d_{Ni}$ (in the studied
range). The mean-square-root roughness of all films is $R_q\simeq
1$ nm, although few spikes up to $\sim \pm 4$ nm can be seen in
all cases. Probably because of that, we could not obtain reliable
data for junctions with $d_{Ni}<5$ nm, which were usually shorted
and did not exhibit JJ behavior. As can be seen from Fig.
\ref{fig:figS1}, the overall roughness of the junction S/F
interface is determined by the roughness of the bottom Nb layer
and is $R_q \simeq 1$ nm for all studied JJ's.

\subsection{Appendix B. Summary of junction characteristics}

\begin{figure}[t]
    \centering
    \includegraphics[width=0.4\textwidth]{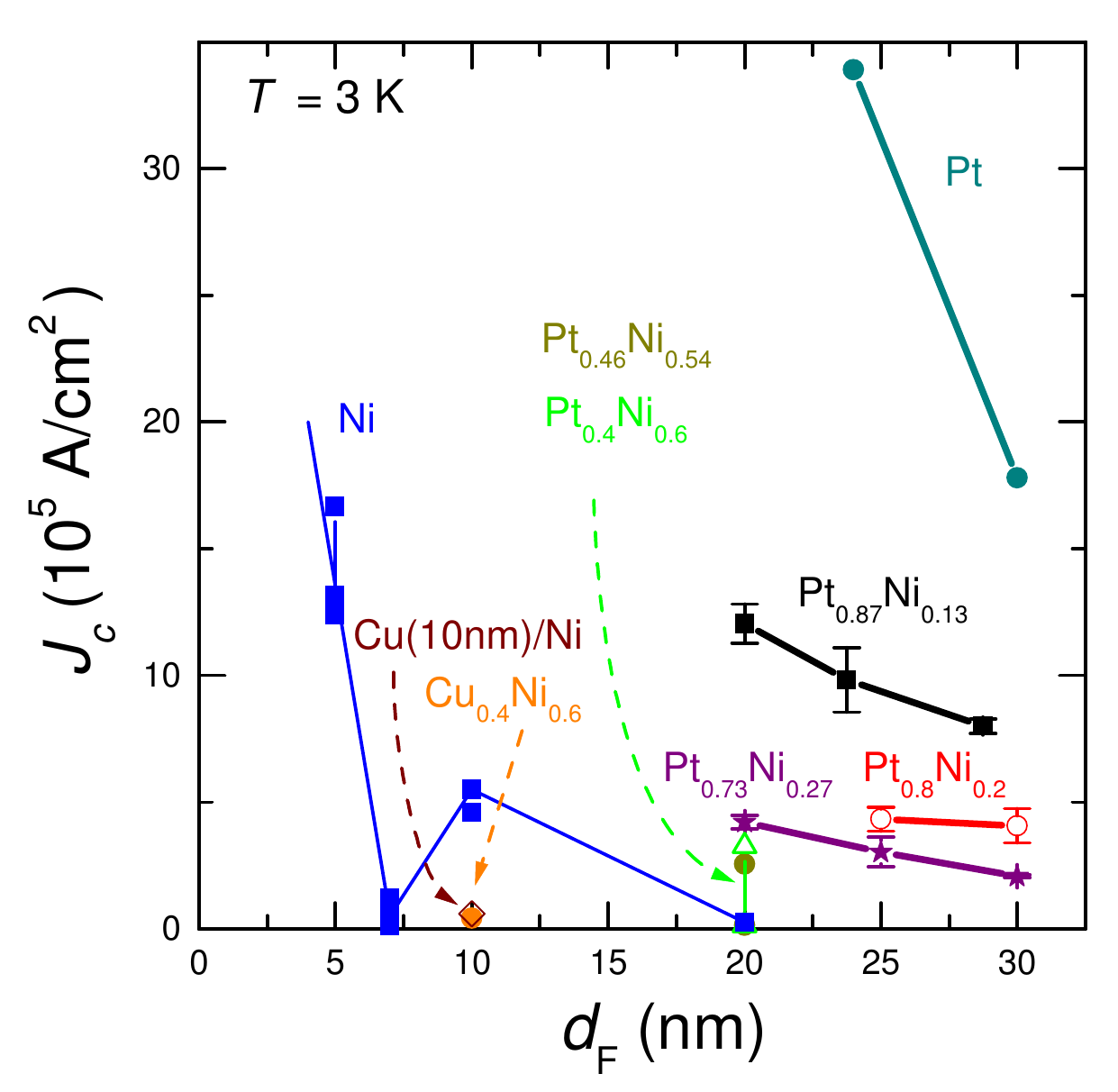}
    \caption{(color online). A summary of measured critical current densities for different junctions at $T=3$ K versus the interlayer thickness.
}
    \label{fig:figS2}
\end{figure}

Tables I-III represent characteristics of all types of studied
junctions. Figure \ref{fig:figS2} summarizes measured critical
current densities at $T=3$ K for JJ's with different interlayer
composition and thickness, studied in this work. $J_c$ decreases
both with increasing Ni concentration and interlayer thickness.
For Nb/Ni/Nb JJ's (blue) we have sufficient samples to observe the
non-monotonous dependence $J_c$ vs. $d_F$ due to $0-\pi$
transitions
\cite{Ryazanov_2001,Kontos_2002,Oboznov_2006,Robinson_2007}. The
blue line, connecting points for Ni-JJ's, however, is drawn solely
for the easiness of identification of the data points and does not
reflect the anticipated $J_c(d_F)$ dependence, which should
oscillate at a much shorter scale $\xi_{F2}(Ni) \sim 1$ nm
\cite{Robinson_2007}.

Our conclusions are based on the overall analysis of SFS
junctions, as listed in Tables I and III. In particular we want to
note that the small $J_c$ value for Ni(20nm) JJ in Fig.
\ref{fig:figS2} is the consequence of the much lower onset
temperature $T^*<4$ K for this JJ, as shown in Fig. \ref{fig:fig3}
(b). The data in Fig. \ref{fig:figS2} is shown for $T=3$ K. This
temperature was chosen because we have data for this $T$ for all
junctions.
The low $T^*$ of Ni-JJ results in the misleadingly low $J_c$(3K),
as can be seen from Fig. \ref{fig:fig1} (f). More appropriate
comparison should be done for $T\ll T^*$. Such data is listed in
Tables I and III and is consistent with our conclusion. One should
also keep in mind the oscillatory dependence of $J_c$ on $d_F$
with a nm-scale period of oscillations. Since the periods are
different for different ferromagnets, it becomes impossible to
make a conclusion by comparing just two JJ's with a fixed $d_F$.
For the same reason we do not claim that $J_c$ in Ni is always
larger than in an alloy (which can not be true due to different
oscillatory dependencies of the two). Also, because of that we can
not make an estimation of decay lengths for our SFS junction.
Somewhat reliable decay length estimation from the data in Fig.
\ref{fig:figS2} could be made only for non-magnetic alloys with
$x<0.4$ (see e.g. black and violet lines in Fig. \ref{fig:figS2}
and Table-II) because such JJ's should not exhibit oscillations.

\begin{table*}
\caption{Parameters of junctions with Ni, Cu/Ni and CuNi
interlayers: $d_F$ is the thickness of F-interlayer; the size
defines junction area $A=L_x\times L_y$; $R_n$ is the normal
resistance of the junction; $\rho_n = R_n A / d$ is junction
resistivity, for junctions with Cu/Ni bilayer $d=d_{Cu}+d_{Ni}=20$
nm, for the rest $d=d_F$; $T$ is the temperature; $I_c$ is the
maximum critical current, $J_c=I_c/A$ is the critical current
density; $I_c R_n$ is the characteristic voltage. For some
junctions values of $I_c$, $J_c$ and $I_c R_n$ at different $T$
are provided. }
\begin{ruledtabular}
\begin{tabular}{lccccccccc}
Interlayer& $d_F$ & Size & $R_n$ & $R_n A$ & $\rho_n$ & $T$ & $I_c$ & $J_c$ & $I_c R_n$ \\

 & (nm) & (nm$^2$) & (m$\Omega$) & ($10^{-10}~\Omega$cm$^2$) & ($10^{-4}~\Omega$cm) & (K) & ($\mu$A) & ($10^4$ (A/cm$^2$) & ($\mu$V) \\
\hline

Ni & 5 & 855$\times$160 & 46.8 & 0.64 & 1.28 & 6.2 & 378 & 27.7 & 17.7 \\

Ni & 5 & 942$\times$237 & 28.3 & 0.632 & 1.26 & 5.2 & 1110 & 49.7 & 31.4 \\

Ni & 5 & 896$\times$164 & 31 & 0.456 & 0.911 & 3 & 2800 & 124 & 86.8 \\
 & &  &  &  &  & 5.5 & 760 & 51.7 & 23.6 \\
\hline

Ni & 7 & 1100$\times$220 & 39.9 & 0.966 & 1.38 & 2 & 30 & 1.24 & 1.2 \\

Ni & 7 & 1380$\times$220 & 21.7 & 0.66 & 0.943 & 2 & 67 & 2.21 & 1.45 \\

Ni & 7 & 950$\times$300 & 32 & 0.912 & 1.3 & 3.5 & 100 & 3.51 & 3.2 \\

Ni & 7 & 750$\times$220 & 68 & 1.12 & 1.6 & 2 & 262 & 15.88 & 17.8 \\

Ni & 7 & 1000$\times$220 & 46 & 1.01 & 1.44 & 3 & 240 & 10.9 & 11.04 \\

\hline

Ni & 10 & 865$\times$165 & 52 & 0.742 & 0.742 & 5.8 & 324 & 22.7 & 16.8 \\

Ni & 10 & 926$\times$250 & 30.5 & 0.706 & 0.706 & 6.5 & 311 & 13.4 & 9.5 \\

Ni & 10 & 925$\times$250 & 29 & 0.671 & 0.671 & 4.5 & 600 & 26 & 17.4 \\
 &  &  &  &  &  & 2 & 1800 & 77.8 & 52.2 \\
\hline

Cu(10 nm)/Ni & 10 & 800$\times$275 & 31.5 & 0.693 & 0.347 & 0.49 & 100 & 4.55 & 3.15 \\

Cu(10 nm)/Ni & 10 & 250$\times$200 & 158 & 0.79 & 0.395 & 0.5 & 8.5 & 1.7 & 1.34 \\

Cu(10 nm)/Ni & 10 & 700$\times$160 & 53 & 0.594 & 0.297 & 1.8 & 13.4 & 1.2 & 0.71 \\

Cu(10 nm)/Ni & 10 & 700$\times$300 & 29.15 & 0.612 & 0.306 & 1.8 & 215 & 10.2 & 6.29 \\

Cu(10 nm)/Ni & 10 & 814$\times$250 & 33 & 0.6716 & 0.3358 & 2.86 & 95 & 4.67 & 3.14 \\

Cu(10 nm)/Ni & 10 & 650$\times$250 & 47.5 & 0.772 & 0.386 & 0.37 & 57.5 & 3.54 & 2.73 \\

Cu(10 nm)/Ni & 10 & 800$\times$175 & 53 & 0.742 & 0.371 & 0.37 & 180 & 12.86 & 9.54 \\

\hline

Cu$_{0.4}$Ni$_{0.6}$ & 10 & 730$\times$230 & 83.5 & 1.41 & 1.41 & 0.56 & 195 & 11.6 & 16.19 \\
\hline

Ni & 20 & 1000$\times$250 & 14 & 0.35 & 0.175 & 0.4 & 500 & 20 & 7.0 \\

\end{tabular}
\end{ruledtabular}
\end{table*}

\begin{table*}
\caption{Parameters of junctions with Pt and paramagnetic PtNi
interlayers. }
\begin{ruledtabular}
\begin{tabular}{lccccccccc}
Interlayer& $d_F$ & Size & $R_n$ & $R_n A$ & $\rho_n$ & $T$ & $I_c$ & $J_c$ & $I_c R_n$ \\

 & (nm) & (nm$^2$) & (m$\Omega$) & ($10^{-10}~\Omega$cm$^2$) & ($10^{-4}~\Omega$cm) & (K) & ($\mu$A) & ($10^4$ (A/cm$^2$) & ($\mu$V) \\
\hline

Pt & 23.75 & 207$\times$104 & 340 & 0.732 & 0.31 & 1.8 & 1400 & 651 & 476 \\

 &  &  &  &  &  & 3.2 & 762 & 354 & 259 \\

Pt & 23.75 & 274$\times$113 & 220 & 0.68 & 0.287 & 1.8 & 1700 & 548 & 374 \\

Pt & 25 & 180$\times$90 & 680 & 1.1 & 0.44 & 2.5 & 160 & 98.7 & 108.8 \\

Pt & 30 & 106$\times$106 & 710 & 0.8 & 0.27 & 3.2 & 200 & 178.6 & 142 \\

Pt & 30 & 170$\times$88 & 500 & 0.75 & 0.25 & 3.2 & 260 & 173.3 & 130 \\

Pt & 30 & 117$\times$88 & 780 & 0.8 & 0.27 & 3.2 & 156 & 151.5 & 121.7 \\
\hline

Pt$_{0.87}$Ni$_{0.13}$ & 20 & 351$\times$85 & 270 & 0.806 & 0.402 & 3.0 & 180 & 60.4 & 67.2 \\

Pt$_{0.87}$Ni$_{0.13}$ & 20 & 308$\times$128 & 160 & 0.631 & 0.315 & 3.1 & 430 & 109 & 68.8 \\

Pt$_{0.87}$Ni$_{0.13}$ & 20 & 330$\times$139 & 133 & 0.61 & 0.305 & 3.1 & 570 & 124 & 75.8 \\

Pt$_{0.87}$Ni$_{0.13}$ & 20 & 372$\times$130 & 125 & 0.605 & 0.303 & 3.2 & 610 & 126 & 76.3 \\

Pt$_{0.87}$Ni$_{0.13}$ & 20 & 340$\times$122 & 138 & 0.572 & 0.286 & 3.2 & 510 & 122.9 & 70.4 \\

Pt$_{0.87}$Ni$_{0.13}$ & 23.75 & 226$\times$222 & 146 & 0.733 & 0.38 & 3.0 & 460 & 91.7 & 67.2 \\

Pt$_{0.87}$Ni$_{0.13}$ & 23.75 & 228$\times$175 & 187 & 0.746 & 0.314 & 3.0 & 360 & 90.2 & 67.3 \\

Pt$_{0.87}$Ni$_{0.13}$ & 23.75 & 235$\times$192 & 133 & 0.60 & 0.253 & 3.0 & 510 & 113 & 67.8 \\

Pt$_{0.87}$Ni$_{0.13}$ & 28.75 & 237$\times$134 & 190 & 0.603 & 0.21 & 2.8 & 240 & 76 & 45.6 \\

Pt$_{0.87}$Ni$_{0.13}$ & 28.75 & 218$\times$180 & 210 & 0.824 & 0.286 & 2.8 & 300 & 76.5 & 63 \\

Pt$_{0.87}$Ni$_{0.13}$ & 30 & 110$\times$100 & 1180 & 1.298 & 0.432 & 2.76 & 28 & 25.4 & 33.0 \\

Pt$_{0.87}$Ni$_{0.13}$ & 30 & 140$\times$120 & 660 & 1.109 & 0.37 & 2.7 & 65 & 38.7 & 42.9 \\

\hline

Pt$_{0.8}$Ni$_{0.2}$ & 20 & 197$\times$144 & 180 & 0.511 & 0.255 & 2.9 & 230 & 81.3 & 41.4 \\

Pt$_{0.8}$Ni$_{0.2}$ & 20 & 229$\times$144 & 210 & 0.693 & 0.345 & 2.9 & 300 & 91.2 & 63 \\

Pt$_{0.8}$Ni$_{0.2}$ & 25 & 287$\times$106 & 302 & 0.919 & 0.367 & 3.0 & 120 & 39.5 & 36.2 \\

Pt$_{0.8}$Ni$_{0.2}$ & 25 & 277$\times$128 & 248 & 0.879 & 0.352 & 3.0 & 155 & 43.7 & 38.4 \\

Pt$_{0.8}$Ni$_{0.2}$ & 25 & 170$\times$106 & 462 & 0.833 & 0.333 & 3.0 & 80 & 44.4 & 37 \\

Pt$_{0.8}$Ni$_{0.2}$ & 25 & 287$\times$106 & 307 & 0.934 & 0.373 & 3.0 & 125 & 41.1 & 38.4 \\

Pt$_{0.8}$Ni$_{0.2}$ & 25 & 319$\times$64 & 390 & 0.796 & 0.318 & 3.0 & 80 & 39.2 & 31.2 \\


Pt$_{0.8}$Ni$_{0.2}$ & 30 & 319$\times$106 & 237 & 0.801 & 0.267 & 3.1 & 120 & 35.5 & 28.4 \\

Pt$_{0.8}$Ni$_{0.2}$ & 30 & 266$\times$128 & 228 & 0.776 & 0.258 & 3.1 & 130 & 38.2 & 29.6 \\

Pt$_{0.8}$Ni$_{0.2}$ & 30 & 319$\times$117 & 181 & 0.676 & 0.225 & 3.1 & 180 & 48.3 & 32.6 \\
\hline

Pt$_{0.73}$Ni$_{0.27}$ & 20 & 210$\times$120 & 415 & 1.046 & 0.523 & 3.2 & 101.5 & 40.3 & 42.1 \\

Pt$_{0.73}$Ni$_{0.27}$ & 20 & 210$\times$170 & 293 & 1.046 & 0.523 & 3.2 & 157 & 44 & 46 \\

Pt$_{0.73}$Ni$_{0.27}$ & 20 & 212$\times$90 & 530 & 1.011 & 0.506 & 3.2 & 72 & 37.7 & 38.2 \\

Pt$_{0.73}$Ni$_{0.27}$ & 20 & 210$\times$190 & 245 & 0.978 & 0.489 & 3.2 & 179 & 44.9 & 43.9 \\

Pt$_{0.73}$Ni$_{0.27}$ & 20 & 202$\times$140 & 360 & 1.02 & 0.509 & 3.2 & 120 & 42.4 & 43.2 \\

Pt$_{0.73}$Ni$_{0.27}$ & 20 & 180$\times$175 & 335 & 1.05 & 0.528 & 3.2 & 137 & 43.5 & 45.9 \\

Pt$_{0.73}$Ni$_{0.27}$ & 20 & 175$\times$140 & 300 & 0.735 & 0.368 & 3.2 & 130 & 53.1 & 39 \\

Pt$_{0.73}$Ni$_{0.27}$ & 20 & 255$\times$96 & 320 & 0.783 & 0.392 & 3.2 & 136 & 55.5 & 43.5 \\

Pt$_{0.73}$Ni$_{0.27}$ & 25 & 158$\times$149 & 375 & 0.883 & 0.353 & 3.1 & 61.5 & 26.2 & 23.1 \\

Pt$_{0.73}$Ni$_{0.27}$ & 25 & 175$\times$123 & 330 & 0.71 & 0.284 & 3.1 & 74 & 34.4 & 24.4 \\

Pt$_{0.73}$Ni$_{0.27}$ & 30 & 266$\times$193 & 203 & 1.04 & 0.347 & 3.1 & 104 & 20.3 & 23.9 \\

Pt$_{0.73}$Ni$_{0.27}$ & 30 & 266$\times$167 & 218 & 0.968 & 0.323 & 3.1 & 95 & 21.4 & 20.7 \\

Pt$_{0.73}$Ni$_{0.27}$ & 30 & 256$\times$140 & 278 & 0.996 & 0.332 & 3.1 & 73 & 20.4 & 20.3 \\

\end{tabular}
\end{ruledtabular}
\end{table*}

\begin{table*}
\caption{Parameters of junctions with ferromagnetic PtNi
interlayers. }
\begin{ruledtabular}
\begin{tabular}{lccccccccc}
Interlayer& $d_F$ & Size & $R_n$ & $R_n A$ & $\rho_n$ & $T$ & $I_c$ & $J_c$ & $I_c R_n$ \\

 & (nm) & (nm$^2$) & (m$\Omega$) & ($10^{-10}~\Omega$cm$^2$) & ($10^{-4}~\Omega$cm) & (K) & ($\mu$A) & ($10^4$ (A/cm$^2$) & ($\mu$V) \\
\hline

Pt$_{0.6}$Ni$_{0.4}$ & 25 & 630$\times$230 & 50 & 0.725 & 0.29 & 2.2 & 160 & 11 & 8 \\

Pt$_{0.6}$Ni$_{0.4}$ & 25 & 770$\times$320 & 50 & 1.23 & 0.493 & 2.2 & 63 & 2.6 & 3.2 \\
Pt$_{0.6}$Ni$_{0.4}$ & 25 & 770$\times$320 & 220 & 5.42 & 2.17 & 2.2 & 2070 & 84 & 455$^*$ \\

Pt$_{0.6}$Ni$_{0.4}$ & 30 & 300$\times$165 & 2000 & 9.9 & 3.3 & 3.0 & 140 & 28.3 & 280$^*$ \\

Pt$_{0.6}$Ni$_{0.4}$ & 30 & 380$\times$180 & 1030 & 5.19 & 1.73 & 3.0 & 7.5 & 1.5 & 7.7 \\

Pt$_{0.6}$Ni$_{0.4}$ & 30 & 260$\times$110 & 2200 & 6.29 & 2.09 & 3.0 & 18.3 & 6.4 & 40.3 \\

Pt$_{0.6}$Ni$_{0.4}$ & 30 & 640$\times$225 & 120 & 1.73 & 0.58 & 3.3 & 0 & 0 & 0 \\
Pt$_{0.6}$Ni$_{0.4}$ & 30 & 640$\times$225 & 300 & 4.32 & 1.44 & 2.8 & 110 & 7.6 & 33 \\
\hline

Pt$_{0.46}$Ni$_{0.54}$ & 20 & 800$\times$225 & 50 & 0.9 & 0.45 & 1.8 & 63 & 3.5 & 3.2 \\
 &  &  &  &  &  & 3.0 & 28 & 1.56 & 1.4 \\

Pt$_{0.46}$Ni$_{0.54}$ & 20 & 1140$\times$230 & 120 & 3.15 & 1.57 & 1.8 & 2100 & 80 & 252$^*$ \\
 &  &  &  &  &  & 3.0 & 800 & 30.5 & 96 \\

Pt$_{0.46}$Ni$_{0.54}$ & 20 & 1140$\times$380 & 22 & 0.95 & 0.477 & 1.8 & 92 & 2.1 & 2.0 \\
\hline

Pt$_{0.4}$Ni$_{0.6}$ & 20 & 1200$\times$300 & 90 & 3.24 & 1.62 & 1.8 & 3490 & 96.9 & 314.1$^*$ \\
 & & & & & & 3.0 & 1480 & 41.2 & 133.2 \\

Pt$_{0.4}$Ni$_{0.6}$ & 20 & 1050$\times$420 & 25 & 1.1 & 0.55 & 1.8 & 136 & 3.08 & 3.4 \\
 & & & & & & 3.0 & 62 & 1.41 & 1.6 \\

Pt$_{0.4}$Ni$_{0.6}$ & 25 & 630$\times$340 & 51 & 1.09 & 0.44 & 0.4 & 81 & 3.8 & 4.1 \\
Pt$_{0.4}$Ni$_{0.6}$ & 25 & 640$\times$340 & 53 & 1.15 & 0.46 & 0.4 & 73 & 3.4 & 3.9 \\
Pt$_{0.4}$Ni$_{0.6}$ & 25 & 670$\times$310 & 53 & 1.10 & 0.44 & 0.4 & 58 & 2.8 & 3.1 \\
Pt$_{0.4}$Ni$_{0.6}$ & 25 & 510$\times$310 & 73 & 1.16 & 0.46 & 0.4 & 73 & 4.6 & 5.3 \\
Pt$_{0.4}$Ni$_{0.6}$ & 25 & 550$\times$330 & 71 & 1.29 & 0.52 & 0.4 & 91 & 5.0 & 6.5 \\
Pt$_{0.4}$Ni$_{0.6}$ & 25 & 460$\times$300 & 86 & 1.19 & 0.47 & 0.4 & 110 & 8.0 & 9.5 \\
Pt$_{0.4}$Ni$_{0.6}$ & 25 & 560$\times$310 & 69 & 1.2 & 0.48 & 0.4 & 49 & 2.8 & 3.4 \\

\hline

Pt$_{0.33}$Ni$_{0.67}$ & 20 & 350$\times$190 & 160 & 1.064 & 0.53 & 2.0 & 0 & 0 & 0 \\

Pt$_{0.33}$Ni$_{0.67}$ & 20 & 350$\times$120 & 260 & 1.092 & 0.55 & 2.0 & 0 & 0 & 0 \\

Pt$_{0.33}$Ni$_{0.67}$ & 20 & 290$\times$182 & 200 & 1.056 & 0.53 & 2.0 & 0 & 0 & 0 \\

Pt$_{0.33}$Ni$_{0.67}$ & 20 & 410$\times$180 & 160 & 1.18 & 0.59 & 2.0 & 0 & 0 & 0 \\

Pt$_{0.33}$Ni$_{0.67}$ & 20 & 490$\times$160 & 130 & 1.02 & 0.51 & 2.0 & 0 & 0 & 0 \\

\hline

Ni & 20 & 1000$\times$250 & 14 & 0.35 & 0.175 & 0.4 & 500 & 20 & 7.0 \\
 & & & & & & 2.0 & 188 & 7.5 & 2.6 \\
 & & & & & & 3.0 & 57 & 2.3 & 0.8 \\

\end{tabular}
$^*$ The very large $I_c R_n$ values are not confident because the
corresponding large $I_c$ is comparable to the onset of the
flux-flow phenomenon in Nb electrodes. This leads to the
non-linear $I$-$V$'s at large bias and makes it difficult to
correctly estimate $R_n$.

\end{ruledtabular}
\end{table*}

\subsection{Appendix C. Properties of Nb/Pt$_{1-x}$Ni$_x$/Nb
junctions }

\begin{figure*}[t]
    \centering
    \includegraphics[width=0.99\textwidth]{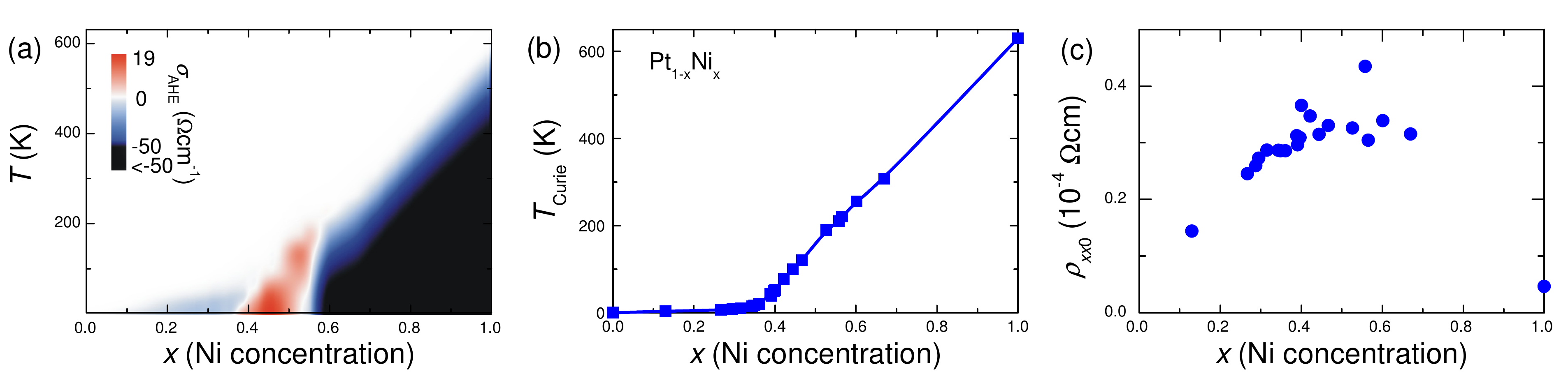}
    \caption{(color online). Characteristics of Pt$_{1-x}$Ni$_x$ thin
    films. (a) Anomalous Hall effect conductivity versus temperature and Ni concentration. (b) Curie temperature obtained from the AHE
    data. Note that the magnetic quantum phase transition with $T_{Curie}\simeq
    0$ occurs at $x_c\simeq 0.4$.
    (c) Residual in-plane resistivity of films. Maxima of $\rho_{xx}$ are observed at $x\simeq 0.4$ and 0.6, corresponding
    to the sign-reversal points of the AHE in panel (a). Data from Refs. \cite{Golod_2011,Golod_2013b}.
}
    \label{fig:figS3}
\end{figure*}

Group-10 elements Pt and Ni are well intermixed with each other
and can form solid solutions at arbitrary proportions
\cite{Dahmani_1985,Besnus_1972,Hizi_2019}. PtNi alloys should be
fairly uniform, contrary to CuNi alloys which are prone to phase
segregation and formation of Ni clusters. Therefore, we have
chosen this alloy for detailed analysis of variation of properties
of SFS junction with the strength of F-interlayer. For that we
made a series of Nb/Pt$_{1-x}$Ni$_x$/Nb JJ's with different Ni
concentrations $x=0-1$. Composition of Pt$_{1-x}$Ni$_x$ films was
estimated using energy-dispersive X-ray spectroscopy
\cite{Golod_2011}.

Figures \ref{fig:figS3} (a) and (b) summarize magnetic properties
of thin Pt$_{1-x}$Ni$_x$ films (35-45 nm thick) obtained in
earlier works \cite{Golod_2011,Golod_2013b}. Fig. \ref{fig:figS3}
(a) shows the anomalous Hall effect (AHE) conductivity
$\sigma_{xy}$ at the $T$-$x$ phase diagram. The AHE indicates
appearance of the ferromagnetic state \cite{Nagaosa_2010}. Fig.
\ref{fig:figS3} (b) shows the Curie temperature extracted from
Hall measurements. It is seen that ferromagnetism in
Pt$_{1-x}$Ni$_x$ thin films appears at $x>0.4$, similar to bulk
alloys \cite{Besnus_1972}.

Pt$_{1-x}$Ni$_x$ alloys may form a disordered fcc state ($A1$),
which is presumably dominant in our sputtered films. However there
are also three ordered states Pt$_3$Ni ($L1_2$), PtNi ($L1_0$) and
Ni$_3$Pt ($L1_2$) with centra of stability at $x=0.25$, 0.5 and
0.75, respectively \cite{Dahmani_1985,Hizi_2019}. The most
remarkable feature of the AHE in PtNi films, Fig. \ref{fig:figS3}
(a), is the sign-change of $\sigma_{xy}$ from electron-like to
hole-like at $x \simeq 0.4-0.6$, which coincides with the expected
range of stability of the layered $L1_0$ PtNi compound
\cite{Dahmani_1985,Hizi_2019,Mokrousov_2011}.

Fig. \ref{fig:figS3} (c) represents residual longitudinal
resistivities of the films at $T=2$ K. The $\rho_{xx0}$ increases
upon mixing of Ni and Pt with maximum around $x\simeq 0.5$. This
indicates a progressive shortening of the electronic m.f.p. due to
the growing disorder. The 50-50 mixture has almost an order of
magnitude larger $\rho_{xx0}$ than the pure Ni film $x=1$.
Simultaneously we also see sharp peaks at $x\simeq 0.4$ and 0.6,
which indicates that additional frustrations in the film structure
appears at the boarders between stability regions of the ordered
$L1_2$ and $L1_0$ states.

Since ferromagnetism in Pt$_{1-x}$Ni$_x$ alloy appears at
$x_c\simeq 0.4$, diluted ferromagnets with small $T_{Curie}\sim
10$ K, comparable to $T_c$ of Nb, correspond to an extremely dirty
metallic state. The short electronic m.f.p. leads to a short
$\xi_F$, which leads to a rapid suppression of the proximity
induced superconducting order parameter with increasing $d_F$
\cite{Buzdin2005,Melnikov_2012}. Therefore, as discussed in the
manuscript, SFS junctions with weak disordered ferromagnets may
have small critical current densities despite small exchange
fields.

\begin{figure}[t]
    \centering
    \includegraphics[width=0.45\textwidth]{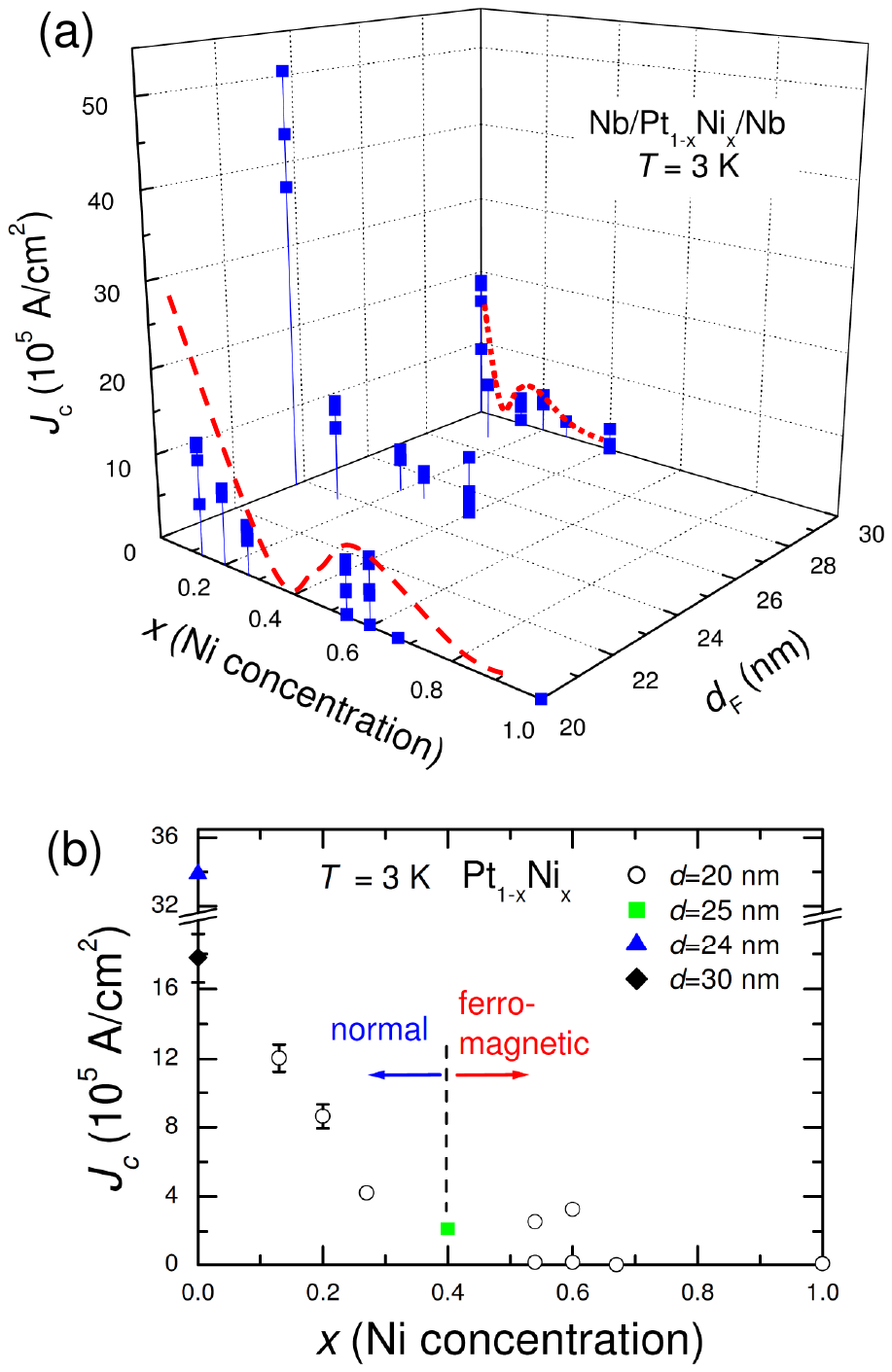}
    \caption{(color online). (a) Three-dimensional plot of critical current densities of
    Nb/Pt$_{1-x}$Ni$_x$/Nb junctions at $T=3$ K as a function of
    Ni concentration and interlayer thickness. (b) Projection of
    the same data to the two-dimensional plot. Red dashed lines in
    (a) indicate possible $0-\pi$ transition for a fixed $d_F$
    upon increasing Ni concentration.
}
    \label{fig:figS4}
\end{figure}

Nb/Pt$_{1-x}$Ni$_x$/Nb junctions with different $x$ and $d_F$ were
fabricated and studied, see Table II. An example of $I_c(H)$
modulation for Nb/PtNi/Nb JJ can be found in Fig. 4 of Ref.
\cite{Golod_2010}.  Figure \ref{fig:figS4} (a) shows measured
$J_c$($T=3$K) for Nb/Pt$_{1-x}$Ni$_x$/Nb junctions versus Ni
concentration and interlayer thickness. Fig. \ref{fig:figS4} (b)
shows projection of this data to the $J_c$-$x$ plane. Generally,
$J_c$ decreases both with increasing $x$ and $d_F$. However, it
decreases non-monotonously. Oscillatory decay of $J_c$ vs. $d_F$
in SFS junctions is well documented and is caused by sequential
0-$\pi$ transitions
\cite{Ryazanov_2001,Oboznov_2006,Robinson_2007,Kontos_2002}.

From Fig. \ref{fig:figS4} it can be seen that for a given $d_F$
the $J_c$ is decaying non-monotonously with increasing Ni
concentration $x$, as indicated by dashed red lines in Fig.
\ref{fig:figS4} (a) for $d_F=20$ and 30 nm. We attribute such
oscillatory behavior to 0-$\pi$ transitions at a given $d_F$ upon
increasing the ferromagnetic exchange energy $E_{ex}$. The
increase of Ni concentration leads to the enhancement of $E_{ex}$,
which leads to the shrinking of $\xi_F$ and cause the 0-$\pi$
transition. As described above, ferromagnetism in Pt$_{1-x}$Ni$_x$
films appears at the critical concentration $x_c\simeq 0.4$. We
observe that the relative spread in $J_c$ values increases in JJ's
with the ferromagnetic interlayer $x>0.4$. Most likely this is
also a consequence of a rapid shrinkage of $\xi_F$ down to about
$1$ nm, comparable to the roughness of our films, see Fig.
\ref{fig:figS1}.

\subsection{Appendix D. Interface resistances in
Nb/Pt$_{1-x}$Ni$_x$/Nb junctions }

\begin{figure}[t]
    \centering
    \includegraphics[width=0.45\textwidth]{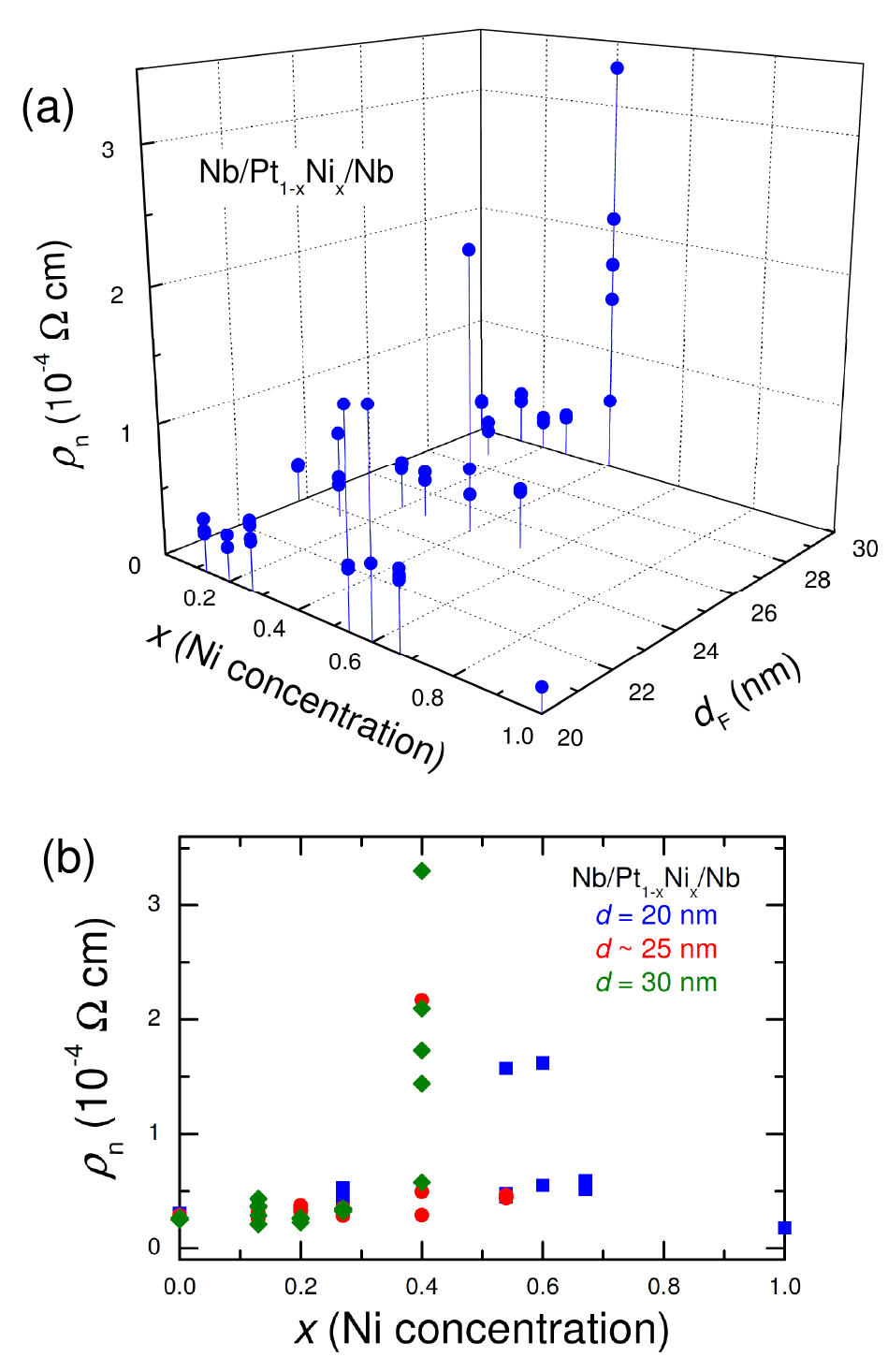}
    \caption{(color online). (a) Three-dimensional plot of normal resistivities of
    Nb/Pt$_{1-x}$Ni$_x$/Nb junctions versus Ni concentration and
    interlayer thickness. (b) Projection of the same data on the
    two-dimensional plot. Note sharp singularities at the critical
    concentration $x_c=0.4$ and a secondary maximum at $x\simeq
    0.6$, corresponding to points of sign-reversal AHE in Fig. S3 (a).
}
    \label{fig:figS5}
\end{figure}

Figure \ref{fig:figS5} (a) shows a 3D plot of measured normal
resistivities, $\rho_n$, of studied Nb/Pt$_{1-x}$Ni$_x$/Nb
junctions versus Ni concentration and interlayer thickness.
Parameters of JJ's are listed in Table III. Here $\rho_n=R_n
A/d_F$, where $A$ is the junction area.
Fig. \ref{fig:figS5} (b) shows the 2D projection of same data. It
is seen that $\rho_n$ greatly increases at $x=0.4-0.6$ which is
correlated with the region with maximal longitudinal resistance of
PtNi films, see Fig. \ref{fig:figS3} (c). Thus frustration and
disorder is directly reflected in junction characteristics.
However, $\rho_n$ exhibits a much larger peaks at the frustration
points $x\simeq 0.4$ and $0.6$, compared to $\rho_{xx0}$.
Especially at the onset of ferromagnetism $x_c=0.4$, where
$\rho_n$ increases by almost an order of magnitude. This indicates
that properties of SFS junctions depend not only on the electronic
disorder (m.f.p.) but also on the magnetic disorder and,
particularly, are affected by quantum fluctuations at the quantum
phase transition reflected by the sign-change of the AHE, Fig.
\ref{fig:figS3} (a).

From the comparison of $\rho_{xx0}$ and $\rho_n$, Figs.
\ref{fig:figS3} (c) and \ref{fig:figS5} (b) it is also seen that
junction resistivity is several times larger than the film
resistivity. This indicates that junction resistances are
dominated by an additional resistance at S/F interfaces due to a
finite interface transparency $\beta <1$. For SNS junctions the
interface transparency is reduced by a mismatch between Fermi
velocities and Fermi surfaces (electronic band structures) of S
and N metals \cite{Krasnov_1992,Buhrman_1998}. For example, the
transparency of Nb/Cu interface was estimated to be $\beta\simeq
0.4$ \cite{Krasnov_1992}. The transparency of S/F interfaces is
further reduced due to spin imbalance, which affects the Andreev
reflection of spin-singlet Cooper pairs from spin-polarized
ferromagnet
\cite{Beenakker_1995,Aarts_1997,Buhrman_1998,Falko_1999,Golubov_1999,Bourgeois_2001,Turek_2002,Tagirov_2003,Yao_2003}.
The values $R_n A \sim 1\times 10^{-10} \Omega$cm$^2$ in our
junctions, see Tables I-III, are comparable to the value
$0.64\times 10^{-10} \Omega$cm$^2$ reported for Nb/Co interfaces
\cite{Yao_2003}. From Table III it can be seen that for
Nb/Pt$_{1-x}$Ni$_x$/Nb JJ's with $x\simeq 0.4$ and 0.6,
corresponding to quantum critical points of vanishing AHE,
$\sigma_{AHE}(T=0)\simeq 0$, see Fig. \ref{fig:figS3} (a) and Ref.
\cite{Golod_2013b}, the $R_n A$ value and, thus, the interface
resistance greatly increases. Simultaneously the critical current
density increases, leading to extraordinary large $I_c R_n$
products of several hundreds of $\mu$V, comparable to that for
SINIS (I-insulator) junctions \cite{Golubov_1995}. The origin of
this phenomenon remains to be understood. So far we can only
speculate that anomalous junction characteristics at these
critical concentrations are related to quantum phase transitions
occurring between ordered $L1_0$ and $L1_2$ phases with different
magnetic properties
\cite{Mokrousov_2011,Dahmani_1985,Hizi_2019,Chen_2011}. For all
our junctions, $R_n$ is dominated by S/F interface resistances,
consistent with earlier reports for other types of S/F interfaces
\cite{Buhrman_1998,Bourgeois_2001,Yao_2003}. Therefore, there are
significant barriers at S/F interfaces, despite the deposition of
SFS trilayers occurred in one run without breaking vacuum.

\subsection{Appendix E. Clarification about extraction of
magnetization curves from AJF analysis}

In derivation of Eq. (1) we assumed that $M_F$ has an in-plane
orientation. Due to the small thickness of F-interlayers, they
have negligibly small demagnetization factors.
In this case the F-layer does not generate magnetic
fields at S/F interfaces and, therefore, does not induce any
additional flux in S-electrodes. This leads to a simple separation
of flux contributions from S-electrodes and F-interlayer,
represented by first and second terms in Eq. (1). Here $B$ in the
first term does not contain $M_F$ (i.e it is not $H+4\pi M_F$) and
differs from $H$ solely due to screening by superconducting
currents and a finite demagnetization factor of S-electrodes, just
like in the non-magnetic junction.
Since for our junctions the total thickness of S-layers $2
d_{Nb}=400$ nm is comparable to junction sizes, the
demagnetization factor of electrodes is non-negligible and the
difference between $B$ and $H$ can be sensible. Nevertheless, this
does not affect the linearity of $\Phi(H)$ curves above the
saturation field because $B\propto H$ in the presence of the
demagnetization effect. Therefore, subtraction of the linear
asymptotics, shown by dashed lines in middle panels of Fig. 2
(a-c), remains unambiguous.

The distance between points in the AJF analysis is determined by
the flux quantization field $\Delta H=\Phi_0/L\Lambda$. It is
smaller for the hard axis orientation of the field, corresponding
to the longest size of the junction $L$. Therefore, AJF curves for
the hard axis, Figs. 2 (c,d), are much more detailed than for the
easy axis orientation, Figs. 2 (a,b). Nevertheless, extraction of
$M_F(H)$ for the hard axis orientation is complicated by two
factors: First, magnetization reversal in the hard axis
orientation occurs initially via coherent rotation of
magnetization (without hysteresis), followed by a small flip, and
continuing coherent rotation towards the saturated state
\cite{Iovan_2002,Ivanov_2013,Iovan_2017,Kapran_2020}. Since the
flip is smaller than $M_{sat}$, it does not allow direct
extraction of $M_{sat}$ from the size of the magnetization jump.
Second, since the length of the electrode $L\sim 1~\mu$m in the
hard axis orientation is much larger than the London penetration
depth of S-electrodes, $\lambda_S \simeq 100$ nm, junctions are
prone to penetration of Abrikosov vortices, which greatly distort
junction characteristics \cite{Golod_2010,Golod_2019}. Therefore,
the field range of our analysis is limited by the range of the
Meissner state.

For the JJ with $d_{Ni}=10$ nm, Fig. 2 (c), the Meissner state
persists up the the saturation state and the straightforward
subtraction of the high-field linear slope from the AJF curves,
shown by the dashed line in the middle panel of Fig. 2 (c),
provides a magnetization loop with the expected saturation at high
fields, as shown in the bottom panel of Fig. 2 (c). It can be seen
that saturation occurs at $H\simeq 1$ kOe, which is consistent
with that obtained using the first-order reversal curves analysis
on the same junction, see Fig. 2 (f) in Ref. \cite{Kapran_2020}.

For the JJ with $d_{Ni}=20$ nm, Fig. 2 (d), the field range is
limited by entrance of Abrikosov vortices. It is smaller than 1
kOe and the saturation is presumably not reached, which does not
allow unambiguous determination of the linear asymptotics. In this
case we have chosen to assume $B=H$ in Eq. (1) and calculate the
first linear term using the definition of the magnetic thickness,
$\Lambda =d_F +
\lambda_{S1}\tanh(d_{S1}/2\lambda_{S1})+\lambda_{S2}\tanh(d_{S2}/2\lambda_{S2})$,
where $d_{S1,2}=200$ nm are is the thicknesses and
$\lambda_{S1,2}=100$ nm are the London penetration depths of the
two Nb-electrodes. The corresponding linear dependence $\Phi_1(H)$
is shown by the dashed line in the middle panel of Fig. 2 (d).
Thus obtained magnetization curve, $M_{Ni}(H_{\perp})$, shown in
the bottom panel of Fig. 2 (d), is in line with the expected
magnetization curve for the hard axis orientation, as discussed
above, and provides a correct value of $M_{sat}$.



\end{document}